\documentclass[a4paper,twocolumn]{esapub} 
\usepackage{natbib}
\usepackage{epsfig}

\title{GAMMA-RAY LINE EMISSION FROM $^{7}$LI AND $^{7}$BE PRODUCTION BY
COSMIC-RAYS}
\author{V. Tatischeff}
\author{J.-P. Thibaud}
\author{J. Kiener}
\affil{Centre de Spectrom\'etrie Nucl\'eaire et de Spectrom\'etrie de Masse,
IN2P3-CNRS, 91405 Orsay, France}
\author{M. Cass\'e}
\affil{Service d'Astrophysique, DAPNIA, DSM, CEA, Ormes des Merisiers, 
91191 Gif-sur-Yvette, France}
\author{E. Vangioni-Flam}
\affil{Institut d'Astrophysique de Paris, 98 bis, Boulevard Arago,
75014 Paris, France}

\def\lsim{\lower.5ex\hbox{$\; \buildrel < \over \sim \;$}}
\def\gsim{\lower.5ex\hbox{$\; \buildrel > \over \sim \;$}}

\begin{document}


\maketitle

\begin{abstract}
We calculate the total $\gamma$-ray line emission at $\sim$450 keV
that accompanies $^7$Li and $^7$Be production by cosmic-ray interactions,
including the delayed line emission at 0.478 MeV from $^7$Be radioactive
decay. We present a new $\gamma$-ray spectroscopic test which has the
potential to give direct information on the nature of the interstellar 
regions into which $^7$Be ions propagate and decay. Finally, we 
evaluate the intensity of the predicted diffuse emission from the central 
radian of the Galaxy.
\end{abstract}

\section{Introduction}

Although it has been known for three decades that cosmic-ray spallation is
important to the origin of the light elements Li, Be and B, neither the
sources of the cosmic-rays nor their interactions with the interstellar
medium are yet well understood (see Vangioni-Flam et al., 2000, for a recent 
review). Future observations with INTEGRAL of the nuclear $\gamma$-ray 
line emission that accompanies the light element production could shed new 
light on this non-thermal nucleosynthesis. The $\gamma$-ray line emission at 
$\sim$450 keV from $^7$Li and $^7$Be production is often detected in 
solar flares and could be important in the 
interstellar medium as well. Its intensity and profile are evaluated in 
Ramaty et al. (1979) and references therein. In this paper, we 
supplement the work of these authors with detailed calculations of the 
delayed line emission at 0.478 MeV that follows the radioactive decay of 
$^7$Be to the first excited state of $^7$Li. 


\section{$^{7}$Be Production}

We considered a steady state, thick target model (e.g. Ramaty et al., 1996)
in which $^{7}$Be is produced by nuclear interactions of accelerated ions 
with a neutral ambient medium of solar composition. We employed an ion 
source spectrum resulting from shock acceleration (e.g. Ramaty et al., 1997)
\begin{equation}
\dot{Q}_i (E_i) \propto {p_i^{-s} \over \beta_i} {\mbox{exp}}(-E_i/E_0)
\end{equation}
where $p_i$, $c\beta_i$ and $E_i$ are, respectively, the momentum, velocity
and kinetic energy per nucleon of the accelerated particle of type $i$, and
$E_0$ is a high-energy cutoff. We used $s$=2, which applies to strong shocks. 
We considered two different compositions for the fast ions: the composition
of the current epoch Galactic cosmic-ray sources (CRS; Ramaty et al., 1997)
and the average composition of the stellar winds and supernova ejecta from
OB associations in the inner Galaxy (OB$_{\mbox{IG}}$; Parizot et al., 1997).

The total cross section for the reaction $^4$He($\alpha$,n)$^7$Be is
from the data compilation of Read \& Viola (1984), with the corrections
suggested by Mercer et al. (1997) to the data of Glagola et al. (1982).
The total cross sections for the CNO spallation reactions by protons and
$\alpha$-particles are also based on
Read \& Viola (1984), but we supplemented the data compilation of these 
authors with the data of Epherre \& Seide (1971) for the $^{14}$N(p,X)$^7$Be
reaction, and of Lafleur et al. (1966) and Inoue \& Tanaka 
(1976) for the $^{16}$O(p,X)$^7$Be reaction. 
We see from Figure~1 that the $\alpha$+$\alpha$
reaction is the main source of $^7$Be for $E_0<100$~MeV/nucleon. We show in
\S 4 that such low values of $E_0$  provide the most favourable cases for 
observing the predicted $\gamma$-ray line emission with INTEGRAL. We 
therefore concentrate in this paper on the $\gamma$-ray line production 
by $\alpha$+$\alpha$ reactions.

\begin{figure}
\centering
\epsfig{file=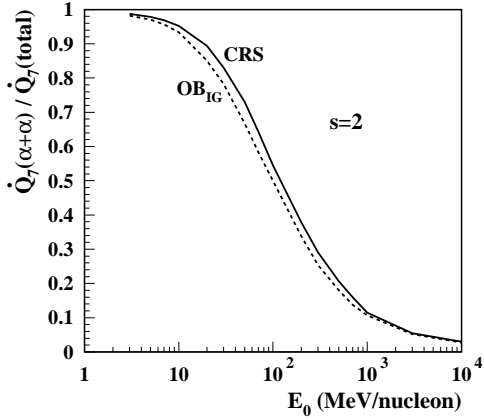,width=0.8\linewidth}
\caption{Ratio of the $^7$Be production rate from the
$^4$He($\alpha$,n)$^7$Be reaction to the total $^7$Be production rate,
as a function of $E_0$ (Eq.~1).}
\end{figure}

The differential cross section for the production of $^7$Be with recoil energy
$E_7$ can be written as
\begin{equation}
{d\sigma_{\alpha \alpha}^* \over dE_7}(E_\alpha,E_7) = \sigma_{\alpha
\alpha}^* (E_\alpha) \cdot {dW_{\alpha \alpha}^* \over d\theta_7^{cm}}
(E_\alpha,\theta_7^{cm}) \cdot {d\theta_7^{cm} \over dE_7}~,
\end{equation}
where $\sigma_{\alpha \alpha}^* (E_\alpha)$ is the total cross section for
the reaction $^4$He($\alpha$,n)$^7$Be, $d\theta_7^{cm} / dE_7$ is a kinematic
factor and $dW_{\alpha \alpha}^* / d\theta_7^{cm}$ is 
the center-of-mass angular distribution of the recoil nuclei, for which we
adopted the energy-dependent analytical formula derived by Murphy et al.
(1988) from available experimental data. 
We show in Figure~2 calculated energy distribution of $^7$Be ions produced
by the $^4$He($\alpha$,n)$^7$Be reaction. 

\begin{figure}
\centering
\epsfig{file=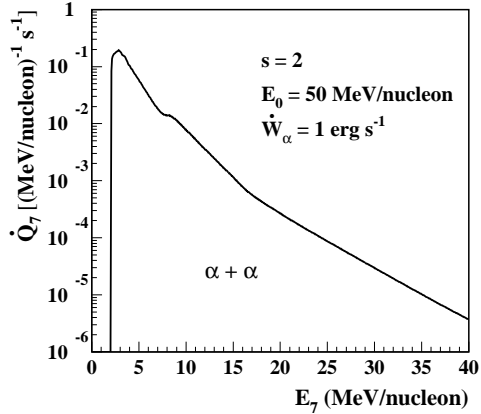,width=0.8\linewidth}
\caption{Thick target differential $^7$Be production rate for the reaction
$^4$He($\alpha$,n)$^7$Be and the $\alpha$-particle source spectrum of Eq.~(1)
with $s$=2 and $E_0$=50 MeV/nucleon. The calculation is normalized such that
the instantaneous energy deposition rate by the accelerated $\alpha$-particles
is 1 erg/s. The energy cutoff at $E_7 \simeq 2$~MeV/nucleon is due to the
reaction kinematics.}
\end{figure}

\section{Gamma-Ray Line Emission From $^7$Be Decay}

$^{7}$Be is a radioisotope which decays by nuclear electron capture
only, with a measured half-life for atoms
$\tau_{1/2}$=53.29 days (Ajzenberg-Selove, 1988). It decays with 10.35~\%
probability to the first excited state of $^{7}$Li, thus producing $\gamma$-ray
line emission at 0.478 MeV. The decay rate depends on the $^{7}$Be
atomic charge and a fully ionized ion can only decay by
highly improbable nuclear electron capture from continuum states. Thus, its
half-life in the interstellar medium is very long, 
$\tau_{1/2}^{(0)}$\gsim10$^{20}$ years (see Bahcall, 1962, eq.~20), so that 
it is stable as compared with the age of the Universe. The decay rates, 
$\lambda^{(n)}$, for $^{7}$Be ions with $n$ bound electrons are given by
\begin{eqnarray}
\lambda^{(4)} & = & {{\rm ln}2 \over \tau_{1/2}} 
               =  \lambda_K + \lambda_L 
	       =  \lambda_K (1 + f_{LK}) \\
\lambda^{(3)} & = & {\lambda^{(4)} \over 1 +
f_{LK}} \left( 1+{f_{LK} \over
2} \right)
\left( \frac{Z - 2{c_K}^\prime}{Z - c_L - 2{c_K}^\prime} \right)^2 \\
\lambda^{(2)} & = & {\lambda^{(4)} \over 1 + f_{LK}} \\
\lambda^{(1)} & = & {\lambda^{(4)} \over 2(1 + f_{LK})} \left( \frac{Z}{Z -
c_K} \right)^2~.
\end{eqnarray}
Here $\lambda_K$ and $\lambda_L$ are the decay rates from the two 1s and
the two 2s electrons, respectively; $f_{LK}$=$\lambda_L / \lambda_K$=3.31~\% 
(Hartree \& Hartree, 1935); $Z$=4 is the $^{7}$Be nuclear charge; and the
shielding constants $c_K$=0.3, ${c_K}^\prime$=0.85 and $c_L$=0.35 (Slater,
1930) take into account the screening of the nuclear charge by the inner
electrons. 


The recoil velocity of $^7$Be nuclei produced by the $^4$He($\alpha$,n)$^7$Be
reaction (see Figure~2) is greater than the orbital velocity of the two
electrons of the $^4$He target atoms. Electron capture into $^7$Be
atomic states during the nuclear reaction is thus improbable and we assumed
in good approximation that the $^7$Be ions are produced fully ionized. As
they slow down by atomic collisions, they capture one or more bound electrons
and then decay to $^7$Li. Thus, it produces a $\gamma$-ray line at 0.478 MeV,
which profile depends on the energy distributions of the $^7$Be$^{(Z-n)+}$ 
ions in the region where they propagate.

\subsection{$^7$Be Equilibrium Spectra}

In a steady state model, the equilibrium spectra of $^{7}$Be ions with
$n$ electrons, $N_7^{(n)}$, satisfy a set of coupled differential equations
of propagation given by
\begin{eqnarray}
{\partial \over \partial E} (N_7^{(n)} \dot{E}_7^{(n)}) = \sum_{j} n_j c 
\beta_7
{\big [} \sigma_{C;j}^{(n-1)} N_7^{(n-1)} - \sigma_{I;j}^{(n)} N_7^{(n)} 
\nonumber \\ 
+ \sigma_{I;j}^{(n+1)} N_7^{(n+1)} - \sigma_{C;j}^{(n)} N_7^{(n)} {\big ]} 
+ \dot{Q}_7^{(n)} - \lambda^{(n)} N_7^{(n)}~.
\end{eqnarray}
Here, $\dot{E}_7^{(n)}$ is the energy loss rate of $^{7}$Be$^{(Z-n)+}$; 
$n_j$ is the number density of species $j$ in the
ambient medium; $\sigma_{C;j}^{(n)}$ and $\sigma_{I;j}^{(n)}$ are,
respectively, the electron capture (i.e. charge exchange) and ionization
cross sections for $^{7}$Be ions with $n$ electrons interacting with neutral
atoms of type $j$; $\dot{Q}_7^{(n)}$ is the energy spectrum of
$^{7}$Be$^{(Z-n)+}$ ions injected into the propagation region (we assumed
that $\dot{Q}_7^{(n)}=0$ for $n>0$); and $\lambda^{(n)}$ is given by 
eqs.~(3-6). We calculated the charge exchange and ionization cross
sections as in Tatischeff et al. (1998). 

We solve numerically Equation~(7) using fifth-order Runge-Kutta method (Press 
et al., 1992). Boundary conditions are provided by the assumptions that at
sufficiently high energies (in practice, we used $E_7 > 10$~MeV/nucleon), 
$^7$Be ions are fully ionized ($N_7^{(n)}=0$ for $n>0$) and both charge 
exchange and radioactive decay are negligible. Equation~(7) then reduces to
\begin{equation}
{\partial \over \partial E} (N_7^{(0)} \dot{E}_7^{(0)}) = \dot{Q}_7^{(0)}
\end{equation}
which admits the solution (e.g. Parizot \& Lehoucq, 1999)
\begin{equation}
N_7^{(0)}(E_7) = {1 \over \dot{E}_7^{(0)}(E_7)} \int_{E_7}^\infty
\dot{Q}_7^{(0)}(E_7') dE_7'~.
\end{equation}

\begin{figure}
\centering
\epsfig{file=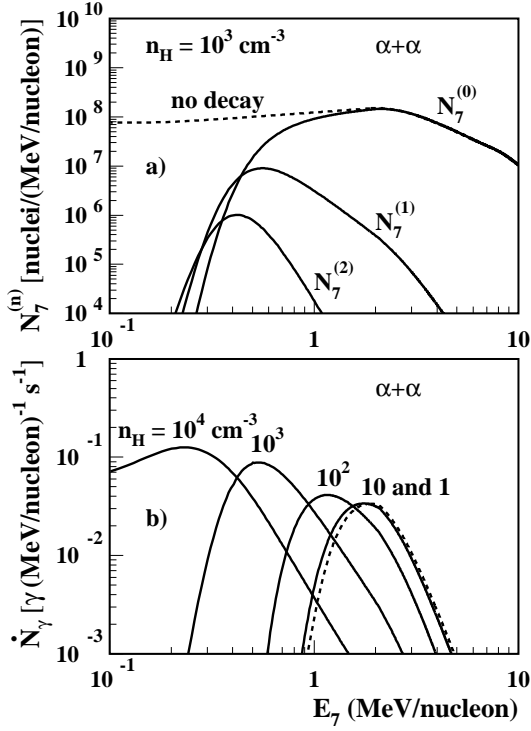,width=0.9\linewidth}
\caption{a) $^7$Be equilibrium spectra in a ambient medium of solar
composition and H density $n_H$=10$^3$~cm$^{-3}$, for the differential
$^7$Be production rate shown in Figure~2. The dashed curve shows the
equilibrium spectrum of test $^7$Be$^{4+}$ particles
which would not capture electron and decay (Eq.~9). (b) Differential
production rates of 0.478 MeV photons from $^7$Be decay as a function
of $^7$Be kinetic energy, for five values of $n_H$ (dashed curve:
$n_H=1$~cm$^{-3}$).}
\end{figure}

Calculated equilibrium spectra are shown in Figure~3a for the differential
$^7$Be production rate of Figure~2 and for an ambient hydrogen density of 
10$^3$ cm$^{-3}$. In that case, $^7$Be ions decay when possessing one or two 
bound electrons, as $\lambda^{(2)}$ is significantly greater than
the characteristic rate for $^7$Be$^{2+}$ ions to capture a third electron.
For simplicity, we assumed that the ambient medium constituents are atomic
species, although such relatively high density is typical of H$_2$ regions.
However, it is a sufficient approximation at this stage, because $^7$Be ions 
capture electrons by colliding mainly with ambient He and heavier elements.

\subsection{Gamma-Ray Line Emission}

The differential $\gamma$-ray line production rate following $^7$Be decay 
is obtained from
\begin{equation}
\dot{N}_\gamma (E_7) = B \sum_{n} \lambda^{(n)} N_7^{(n)} (E_7)~,
\end{equation}
where $B$=10.35~\% is the branching ratio to the first excited state of
$^7$Li. The results are shown in Figure~3b for five values of $n_H$ and for
the differential $^7$Be production rate of Figure~2. We see that for
$n_H>10$~cm$^{-3}$, the differential $\gamma$-ray production rate depends on
the density of the $^7$Be propagation region.
This is not the case for $n_H \leq 10$~cm$^{-3}$, because the $\gamma$-ray 
line is then produced by the decay of $^7$Be$^{3+}$ ions, which do not loose 
significant energy after having captured one electron.

\begin{figure*}
\centering
\epsfig{file=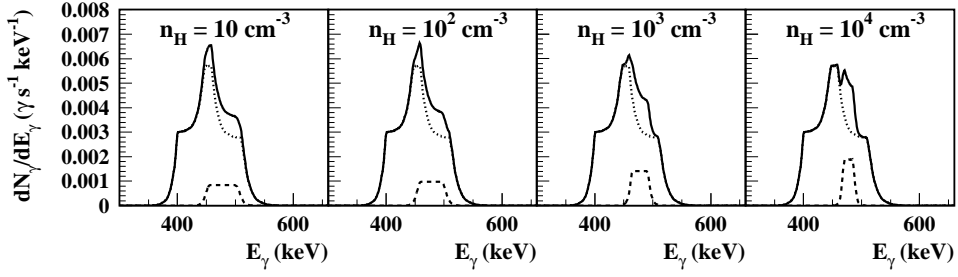,width=0.75\linewidth}
\caption{Calculated $\gamma$-ray line profiles from $\alpha$+$\alpha$ reactions 
in the interstellar medium, including both the prompt lines from the reactions
$^4$He($\alpha$,${\mbox{n}}\gamma_{0.429}$)$^7$Be and
$^4$He($\alpha$,${\mbox{p}}\gamma_{0.478}$)$^7$Li (dotted curves) and the 
delayed line at 0.478 MeV from $^7$Be decay (dashed curves), 
for the same $\alpha$-particle source spectrum as in Figure~2. Solid curves 
show the sum of the two contributions.}
\end{figure*}

The energy of emitted photons is related to the kinetic energy of excited 
$^7$Li* nuclei, $E_{7*}$, by the usual Doppler formula. We assumed the 
$\gamma$-ray emission to be  isotropic in the $^7$Li* rest frame and 
neglected in good approximation the recoil of the $^{7}$Li* nuclei during 
neutrino emission ($E_\nu=0.384$~MeV), so that $E_{7*}=E_7$.
Calculated profiles of the 0.478 MeV line from $^7$Be decay are shown in
Figure~4, together with the prompt $\gamma$-ray line emission from the
reactions $^4$He($\alpha$,${\mbox{n}}\gamma_{0.429}$)$^7$Be and
$^4$He($\alpha$,${\mbox{p}}\gamma_{0.478}$)$^7$Li.
We calculated the latest in the same
steady state, thick target interaction model as for the $^7$Be production (see
\S~2) and from the total and differential cross sections given in Murphy et al.
(1988) and references therein. We assumed the angular distribution of the
accelerated $\alpha$-particles to be isotropic, which leads to relatively
broad, prompt $\gamma$-ray lines ($\delta E/E=17$~\%). These lines blend to
form a broad emission feature (FWHM$\simeq$120 keV) with a narrower 
enhancement at the center. Unlike the prompt emission, the profile of the 
delayed line at 0.478 MeV depends on the ambient medium density, as $^{7}$Be 
ions decay at lower energies in denser propagation regions. Although the 
delayed emission accounts for $\sim$10~\% of the total $\gamma$-ray line 
production, we see that the $\gamma$-ray spectra are significantly modified 
for different values of $n_H$. Thus, the prominent feature at $\sim$460 keV 
has a width of $\sim$50 keV for $n_H=10^4$~cm$^{-3}$, against $\sim$20 keV 
for $n_H=10^2$~cm$^{-3}$. Both SPI and IBIS have potentially
sufficient energy resolution to distinguish between these two emissions.
Therefore, future fine spectroscopic analyses of the emission
profile could in principle allow to better understand the complex 
interactions of low-energy cosmic-rays with dense molecular clouds. 


\section{Predicted Galactic Fluxes}

\begin{figure}
\centering
\epsfig{file=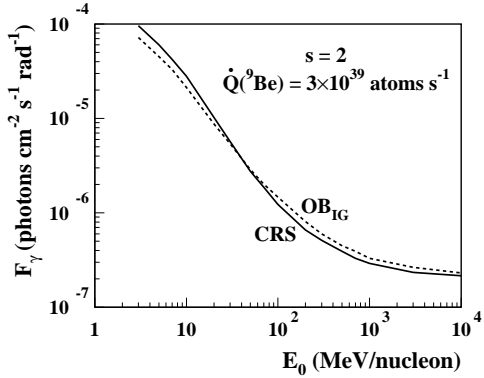,width=0.8\linewidth}
\caption{Predicted fluxes of both prompt and delayed $\gamma$-ray line
emission at $\sim$450 keV from the central radian of the Galaxy.
The calculations are normalized to a Galaxy-wide $^9$Be production
by the fast particles of $3\times10^{39}$~atoms s$^{-1}$ (see text).}
\end{figure}

Ramaty et al. (1997) have shown that the
observation of the quasi linear correlation between Be and Fe abundances in
low metallicity stars implies that the current epoch, Galaxy-wide $^9$Be
production rate is  $\dot{Q}({^9\mbox{Be}})\simeq 3\times10^{39}$~atoms 
s$^{-1}$. This result does not depend on the origin of the $^9$Be-producing 
cosmic-rays, which is not well understood. 
Figure 5 shows predicted $\gamma$-ray line fluxes at $\sim$450 keV (both
prompt and delayed emissions) from the central radian of the Galaxy, assuming
that fast particles with CRS or OB$_{\mbox{IG}}$ composition and with source
spectrum resulting from strong shock acceleration (Eq.~1 with $s$=2) are
responsible for all the current era $^9$Be production. We calculated the
latest as Ramaty et al. (1997) and also used the same spatial model as these
authors. For the prompt $\gamma$-ray line emission resulting from the
CNO spallation reactions, we simply assumed (given the lack of experimental
data) that (i) the productions of excited $^7$Li and $^7$Be nuclei are equal
and (ii) 10~\% of the total $^7$Be production is accompanied by
the decay of its first excited state. This value is based on a recent
measurement of the $^{14}$N($p$,2$\alpha\gamma_{0.429}$)$^7$Be reaction 
cross section (J. Kiener et al., in preparation).

We see that the predicted $\gamma$-ray fluxes strongly increase with
decreasing $E_0$, because of the higher contribution of $\alpha$+$\alpha$
reactions, which do not produce $^9$Be nuclei. However, Ramaty et al. (1997)
have argued from energetic arguments that the Galaxy-wide $^9$Be production
can not be due to low-energy cosmic-rays with $E_0<$ 50~MeV/nucleon.
We thus predict that if $^7$Li and $^9$Be are indeed produced by the same
cosmic-rays, the $\sim$450 keV nuclear line emission from the central radian
of the Galaxy should not exceed $\sim$$3\times10^{-6}$ photons cm$^{-2}$
s$^{-1}$. This Galactic background may be difficult to observe with INTEGRAL.
However, it does not preclude a possible detection of the $\sim$450 keV
line emission from nearby, active regions, such as Cygnus, Vela or Orion.


\end{document}